# A Review of Multi-Objective Deep Learning Speech Denoising Methods

Arian Azarang, *Student Member, IEEE*, Nasser Kehtarnavaz, *Fellow, IEEE*

*Abstract*—This paper presents a review of multi-objective deep learning methods that have been introduced in the literature for speech denoising. After stating an overview of conventional, single objective deep learning, and hybrid or combined conventional and deep learning methods, a review of the mathematical framework of the multi-objective deep learning methods for speech denoising is provided. A representative method from each speech denoising category, whose codes are publicly available, is selected and a comparison is carried out by considering the same public domain dataset and four widely used objective metrics. The comparison results indicate the effectiveness of the multi-objective method compared with the other methods, in particular when the signal-to-noise ratio is low. Possible future improvements that can be achieved are also mentioned.

## I. Introduction

It is well known that noisy environments have a degrading effect on the quality and intelligibility of speech signals. Operating in noisy environments affect the performance of speech processing systems that are normally designed based on clean speech signals [1-16]. Speech denoising involves the reduction or removal of the noisy part of the speech+noise mixture. Speech denoising has been extensively studied in the speech processing literature and have been deployed in a wide range of speech processing systems including speech recognizers [17-18], voice activity detectors [19-21], hearing aids [22-23] and voice over IP (VoIP) [24].

Initial attempts at speech denoising involved applying linear filters to the mixture to reduce or eliminate the noise portion, most prominently Wiener filtering [25], by estimating the noise statistics. In this review, speech denoising methods are placed into two main categories [26-29]: conventional methods, including Wiener filtering, spectral subtraction, and Minimum Mean Square Error (MMSE) methods [30-39], and more recent deep learning-based methods [40-51]. Conventional methods attempt to estimate the statistical attributes in the mixture. A shortcoming of conventional methods is that their effectiveness significantly diminishes in nonstationary noisy environments [52-53]. More recently, deep learning methods have been introduced for speech denoising which are shown to perform better in nonstationary noisy environments compared with conventional methods. Deep learning methods attempt to model the nonlinear relationship between the mixture and the clean speech signals without knowing the noise statistics [54-56]. There have also been speech denoising solutions where conventional and deep learning methods are combined (named hybrid methods) [57-59]. In the hybrid methods, denoised speech signals are obtained by incorporating conventional methods into the deep learning framework. The latest deep learning methods have incorporated not a single objective or loss function to optimize the parameters of a deep neural network but rather a combination of a number of different objective or loss functions (named multi-objective deep learning methods) [60-75].

A number of reviews of speech denoising has previously appeared in [26-29]. This review differs from the previous reviews by focusing on the multi-objective deep learning methods, which are the state-of-the-art speech denoising methods at the time of this writing. For the sake of completeness and for the purpose of laying the mathematical foundation to review the multi-objective deep learning methods, an overview of conventional methods, single objective deep learning methods, and hybrid methods is first stated in Section II. Then, in Section III, the multi-objective deep learning methods are reviewed. A comparison of representative methods in the above speech denoising categories is reported in Section IV followed by mentioning possible future improvements in Section V. Finally, the paper is concluded in Section VI.

## II. Speech Denoising Methods

### A. Conventional Methods

A widely used and one of the earliest conventional methods is Wiener filtering [25]. In general, conventional methods attempt to obtain an estimate of the noise statistics. Another conventional method is spectral subtraction, developed by Boll [76], where the noise in the frequency domain over non-speech activity segments is used to subtract it from the mixture during speech activity segments. MMSE-based [77] and its variations [30-39] is another popular conventional method. As stated earlier, a major shortcoming of these conventional methods is that their performance is adversely impacted in the presence of nonstationary noisy environments.

### B. (Single Objective) Deep Learning Methods

After the initial introduction of deep learning networks by Hinton in 2006 [78], different deep learning configurations have been introduced such as Autoencoders [79], Convolutional Neural Networks (CNNs) [80], and Recurrent Neural Networks (RNNs) [81]. Autoencoders are designed to reconstruct the input at their output. They consist of encoding layers and decoding layers. Encoding layers abstract the input by removing redundant information in the input, while decoding layers reverse the process. CNNs first obtain high

The authors are with the Department of Electrical and Computer Engineering, University of Texas at Dallas, Richardson, TX. Email addresses: {azarang, kehtar}@utdallas.edu

level representations of the input through several convolutional layers which are then distinguished by fully connected layers. RNNs are similar to CNNs but incorporate feedback connections in addition to forward connections.

Single objective deep learning methods can be placed into two major groups: (1) Time-frequency domain methods, e.g. [82-85, 89], in which generally LPS features are used as the input and the target of a deep neural network. Mapping-based (direct mapping) methods, e.g. [51, 84-85], in which the magnitude of LPS features are considered as training samples. Masking-based methods constitute another category in this group, e.g. [86-88], where the target of the network is considered to be an intermediate mask indicating the presence of clean and noise signals. This approach was first introduced in [86] by considering a binary mask that assigned 0 or 1, respectively, to the existence or absence of noise. In [87], a soft mask was considered instead of a binary mask [86], which was shown to be more effective compared to the binary mask. Several improvements to masking-based methods have also appeared. For example, in [90], a phase-sensitive filter was proposed which incorporated the phase information into the estimated mask. In another category in this group, a set of complementary features as well as the masking and mapping-based targets developed for the training of deep neural networks [89] (2) Time-domain methods, e.g. [91-99], in which raw data in the time domain is considered. One of the earliest works in this group was reported in [91], where a fully convolutional neural network was used to map noisy time domain signals to their clean version. Another example is the method discussed in [92] where an end-to-end speech denoising was performed by minimizing the reconstruction error in the time domain.

*C. Hybrid Methods*

Tu et al. introduced a hybrid method in [57] by incorporating conventional methods into the deep learning framework. An LSTM Multi-Style (LSTM-MT) model was trained using LPS as the input and clean LPS and Ideal Ratio Mask (IRM) as the targets. In what follows, the equations for the hybrid methods are briefly mentioned in order to lay the mathematical foundation for the review of the multi-objective methods in the next section.

In the above hybrid method, the denoising process for the $l$-th audio frame is achieved in three steps. The first step, called Approximate Speech Signal Estimation (ASSE), preprocesses the noisy LPS $X(l)$ by considering a suppression rule, resulting in a clean speech estimate $Y(l)$. In the second step, the trained LSTM-MT network uses $Y(l)$ to generate estimates of IRM $M(l)$ and clean speech $\hat{S}(l)$. In the final step, the estimates of IRM $M(l)$ and clean speech $Y(l)$ are utilized to estimate the output speech signal $Z(l)$. The above steps are elaborated below.

The prior and posterior Signal-to-Noise Ratios (SNRs), denoted by $\gamma(k,l)$ and $\xi(k,l)$, respectively, are estimated using the following widely used equations

$$\gamma(k,l) \triangleq \frac{|X(k,l)|^2}{\sigma(k,l)}, \xi(k,l) \triangleq \frac{|S(k,l)|^2}{\sigma(k,l)} \quad (1)$$

where $\sigma(k,l)$ indicates the noise variance in time frame $l$ and frequency bin $k$. Then, the noise reduction or suppression is achieved by the gain $G(k,l)$ as follows

$$G(k,l) = g(\gamma(k,l), \xi(k,l)) \quad (2)$$

where $g(.)$ is a function of prior and posterior SNRs described in [66]. ASSE is then computed using the following equation

$$Y(k,l) = log[\delta M(k,l) + (1-\delta)G(k,l)] + X(k,l) \quad (3)$$

where $\delta$ denotes a weighting parameter.

Next, IRM is computed which is the ratio of the powers of the clean and noisy speech signals according to

$$M_{ref}(k,l) = \frac{|S(k,l)|^2}{|X(k,l)|^2} \quad (4)$$

The objective function for training is the combination of the terms noted below

$$E = \sum_{k,l} \left[ \left(\hat{S}(k,l) - S(k,l)\right)^2 + \left(M(k,l) - M_{ref}(k,l)\right)^2 \right] \quad (5)$$

In another more recent hybrid method in [58], the speech signal $S(k,l)$ is recovered and the noise signal $V(k,l)$ is reduced by applying a linear filter transfer function $H(k,l)$, for example Wiener filter, to the observation $X(k,l)$. The Wiener filter transfer function is given by

$$H(k,l) = \frac{\phi_{SS}(k,l)}{\phi_{SS}(k,l) + \phi_{VV}(k,l)} \quad (6)$$

where $\phi_{VV}(k,l)$ denotes the Power Spectral Density (PSD) of the noise signal expressed as

$$\phi_{VV}(k,l) = E\{V(k,l)V^*(k,l)\} \quad (7)$$

Likewise, the PSD of the desired signal is expressed as

$$\phi_{SS}(k,l) = E\{S(k,l)S^*(k,l)\} = \phi_{XX}(k,l) - \phi_{VV}(k,l) \quad (8)$$

where $\phi_{XX}(k,l) = E\{X(k,l)X^*(k,l)\}$ denotes the PSD of the observed signal. For real-time operation, at time frame $l$, the PSDs of the noise signal and the observed signal can be updated recursively as follows

$$\hat{\phi}_{VV}(k,l) = \hat{\alpha}_V(k,l)\hat{\phi}_{VV}(k,l-1) \\ + (1 - \hat{\alpha}_V(k,l))X(k,l)X^*(k,l) \quad (9)$$

$$\hat{\phi}_{XX}(k,l) = \hat{\alpha}_X(k,l)\hat{\phi}_{XX}(k,l-1) \\ + (1 - \hat{\alpha}_X(k,l))X(k,l)X^*(k,l) \quad (10)$$

where $\hat{\alpha}_V$ and $\hat{\alpha}_X$ denote smoothing weights for the PSDs of the noise and the input mixture, respectively.

The PSD of the noise signal can be estimated based on the past average of power spectral values in a time-varying smoothing manner using a so-called Speech Presence Probability (SPP). However, when noise is nonstationary, or the level of SNR is low, it is quite challenging to acquire an accurate SPP, which limits the tracking capability of the noise estimator in case of time-varying noise.

To address this limitation, a deep learning network is used to replace SPP. This way, conventional and deep learning modules are merged together as part of the same solution. Two-layers of a Gated Recurrent Unit (GRU) and a one-layer of a feedforward network are utilized to construct a deep noise tracking network. A sigmoid output layer follows the GRU generating a vector of 0 and 1 elements denoted by $\alpha_V$, which is used as the SPP of the current frame enabling an adapting smoothing factor. After updating the network parameters, the final desired speech signal is obtained via the following equation

$$\hat{S}(k,l) = \frac{\phi_{XX}(k,l) - \phi_{VV}(k,l)}{\phi_{XX}(k,l)} X(k,l) \quad (11)$$

The following MSE-based magnitude objective function is used to optimize the network parameters

$$J = \frac{1}{L}\sum_{l=0}^{L}\sum_{k=0}^{K}|\hat{S}(k,l) - S(k,l)|^2 \quad (12)$$

where $L$ and $K$ correspond to the total number of frequency bins and time frames, respectively.

A real-time low-complexity hybrid speech denoising method was also discussed in [59] which uses a combination of a conventional method within a RNN network.

### III. Multi-Objective Deep Learning Methods

A multi-objective deep learning method, named LSTM-RNN, was first proposed in [60]. Initially, a Direct Mapping (DM) method is utilized based on a linear output layer and this Minimum Mean Square Error (MMSE) objective function

$$E_{DM} = \sum_{k,l}\left(\hat{S}(k,l) - S(k,l)\right)^2 \quad (13)$$

in which $\hat{S}(k,l)$ and $S(k,l)$ are the estimated and the reference clean LPS features at the (Time-Frequency) T-F bin, respectively. LSTM-IRM is another deep learning method introduced in [87] which is an extension of the Ideal Binary Mask (IBM) method. IRM has been shown to be effective as a soft mask defined as follows

$$M_{ref}(k,l) = \frac{S(k,l)}{S(k,l) + V(k,l)} \quad (14)$$

The corresponding objective function is stated as

$$E_{IRM}(k,l) = \sum_{k,l}\left(\widehat{M}(k,l) - M_{ref}(k,l)\right)^2 \quad (15)$$

As noted earlier $\widehat{M}(k,l)$ and $M_{ref}(k,l)$ are the estimated and reference IRM, respectively.

LSTM-DM would be the right option for speech denoising if the underlying clean speech can be perfectly reconstructed. However, due to data limitations and the local minima aspect of the LSTM-RNN optimization, it would be difficult to learn the relationship of the noisy and clean speech signals in LSTM-DM. As a result, one cannot say which approach is better, in particular in the presence of unseen noise signals, and across different SNR levels.

The idea here is to jointly learn the clean speech and IRM in one single LSTM-RNN using dual outputs, called Multi Task Learning (MTL), whose objective function is expressed as

$$E_{MTL} = \sum_{k,l}\left[\left(\hat{S}(k,l) - S(k,l)\right)^2 \\ + \alpha_{MTL}\left(\widehat{M}(k,l) - M_{ref}(k,l)\right)^2\right] \quad (16)$$

where $\alpha_{MTL}$ denotes a weight for the dual outputs $\hat{S}(k,l)$ and $\widehat{M}(k,l)$. During the denoising stage, the estimated clean speech and IRM can be combined via an averaging operation in the LPS domain as follows

$$\hat{Z}(k,l) = \frac{1}{2}[\hat{S}(k,l) + \log \widehat{M}(k,l) + X(k,l)] \quad (17)$$

Another multi-objective framework was proposed in [61] in order to optimize a joint objective function, involving errors not only for the primary clean LPS features but also errors in the secondary target for continuous features, such as MFCCs, and for categorical information, such as IBM. This joint optimization of different but related targets increases the DNN performance on the prediction of the primary target LPS which is then used to reconstruct the denoised output signal. When a DNN is considered to serve as a mapping function between the noisy and clean LPS features, no assumption is imposed during the training process. On the other hand, in masking-based methods, some constraints, such as the independency of noise and clean speech, need to be imposed.

The multi-objective learning approach considers the loss function of LPS features together with MFCC features and the IBM mask as noted below

$$E = \frac{1}{N}\sum_{l=1}^{N}\frac{\|\hat{S}_l(X_l,X_{l\pm\tau}^{cont},W,b)-S_l\|_2^2}{\|x_n\|_2^2} + \alpha_{cont} * \frac{1}{N}\sum_{l=1}^{N}\frac{\|\hat{S}_l^{cont}(X_{l\pm\tau},X_{l\pm\tau}^{cont},W,b)-S_l^{cont}\|_2^2}{\|X_l^{cont}\|_2^2} + \alpha_{cate} * \frac{1}{N}\sum_{l=1}^{N}\|\hat{S}_l^{cate}(X_{l\pm\tau},X_{l\pm\tau}^{cate},W,b) - S_l^{cate}\|_2^2 \quad (18)$$

where $\hat{S}^{cont}$ and $S^{cont}$ correspond to the estimated and clean features, $\hat{S}^{cate}$ and $S^{cate}$ denote the estimated and target meta category information, $\alpha_{cont}$ and $\alpha_{cate}$ are the weighting coefficients of the second and third terms, respectively. The prediction for the secondary feature is complementary to the primary LPS features. The IBM learning can also improve the clean speech estimation.

Although many improvements have been made by so called Denoising AutoEncoders (DAEs) in speech denoising applications, they often yield speech distortion due to over-smoothing and clipping clean speech using the MSE objective loss. This affects the perceptual speech quality leading to a muffled sound. A perceptron optimized deep denoising autoencoder for single channel speech denoising appeared in [63] which is described next.

The conventional DAE uses the MSE loss function to map the noisy to the clean signal. This loss or error is defined as

$$E = \frac{1}{2}\|S - \hat{S}\|_2^2 \quad (19)$$

where $\hat{S}$ is the output of the DAE in feedforward propagation and $S$ is the clean target mentioned before. The backpropagation of the error is derived according to

$$\frac{\partial E}{\partial \hat{S}} = \frac{\partial}{\partial \hat{S}}\frac{1}{2}(S - \hat{S})^2 = \hat{S} - S \quad (20)$$

The linearity in the computation of the above gradient often leads to the over-smoothing problem typically seen in DAE. This is due to the fact that the penalty for clipping speech is the same as clipping noise as long as the Euclidian distance from the clean target is the same. In this method, to maintain the perceptual quality, it is better to preserve speech segments with the residual error rather than clipping speech segments to remove noise. Hence, a new loss has been designed which considers high penalty against signal removal and preserves the same error for noise removal. This objective function is given by

$$E = \begin{cases} \frac{1}{2}\|S - \hat{S}\|_2^2 & if\ \hat{S} \geq S \\ \frac{1}{2}\|S - \hat{S} + p\|_2^2 & if\ \hat{S} < S \end{cases} \quad (21)$$

where $p$ is a positive scalar denoting the penalty of speech clipping. The gradient is then computed as follows

$$\frac{\partial E}{\partial \hat{S}} = \begin{cases} \hat{S} - S & if\ \hat{S} \geq S \\ \hat{S} - S - p & if\ \hat{S} < S \end{cases} \quad (22)$$

This new objective function is equivalent to the conventional MSE loss when the penalty is set to zero ($p = 0$). The training phase of DAE leads to an estimate of the clean signal while avoiding the removal of the desired signal.

In another study, a perceptually guided speech denoising was proposed via a deep neural network [64]. From the hearing perception perspective, the MSE loss function is not optimal. It helps to incorporate the quality and intelligibility knowledge of speech into the loss function. In this approach, the short-time objective intelligibility measure (STOI) [100] is added to the objective function to optimize speech intelligibility. For this new loss function, the following steps are considered to derive the modified STOI function. Assuming a 16 kHz sampling rate, for each time frame, a 512-point Fast Fourier Transform (FFT) is taken yielding 25 frequency bins. Then, the frequency bins are grouped together to form one-third octave bands. Let $S(k, l)$ and $\hat{S}(k, l)$ denote the STFT representation of the clean and denoised speech signals, respectively. The corresponding frequency bins are grouped into 15 one-third octave bands. Then, the new T-F representation appears as follows

$$S_j(l) = \sqrt{\sum_{k=k_1(j)}^{k_2(j)-1}\|S(k,l)\|_2^2} \quad (23)$$

$$\hat{S}_j(l) = \sqrt{\sum_{k=k_1(j)}^{k_2(j)-1}\|\hat{S}(k,l)\|_2^2} \quad (24)$$

where $j$ corresponds to the index of one-third octave bands, $k_1$ and $k_2$ denote the edges of these bands, and $\|\cdot\|_2$ indicates $L_2$ norm. As a result, the short-term temporal envelope of the clean and denoised signals can be expressed as

$$s_{l,j} = [S_j(l), S_j(l+1), ..., S_j(l+R-1)]^T \quad (25)$$

$$\hat{s}_{l,j} = [\hat{S}_j(l), \hat{S}_j(l+1), ..., \hat{S}_j(l+R-1)]^T \quad (26)$$

Noting that the analysis window is 384ms in length, $R$ is set to 24. Based on the original STOI computation, the following equation is used to normalize and clip the denoised speech signal

$$\bar{s}_{l,j}(i) = \min(\frac{\|s_{l,j}\|_2}{\|\hat{s}_{l,j}\|_2}\hat{s}_{l,j}(i), \left(1 + 10^{-\frac{\beta}{20}}\right)s_{l,j}(i)) \quad (27)$$

where $i = 1,2, ..., N$; $\beta$ controls the lower bound of the Signal to Distortion Ratio (SDR).

The intermediate speech intelligibility measure is computed using the correlation coefficients between the vectors $s_{l,j}$ and $\bar{s}_{l,j}$, namely

$$d_{l,j} = \frac{\left(s_{l,j} - \mu_{s_{l,j}}\right)^T \left(\bar{s}_{l,j} - \mu_{\bar{s}_{l,j}}\right)}{\left\|s_{l,j} - \mu_{s_{l,j}}\right\|_2 \left\|\bar{s}_{l,j} - \mu_{\bar{s}_{l,j}}\right\|_2} \quad (28)$$

in which $\mu(.)$ indicates the mean vector. The modified STOI function is defined as follows

$$d_l = f(S_l^{24}, \hat{S}_l^{24}) = \frac{1}{J}\sum_j d_{l,j} \quad (29)$$

where $S_l^{24}$ and $\hat{S}_l^{24}$ are the 24-frame magnitude spectrum starting from the time frame $m$ of the clean and denoised speech signals, respectively, and $J$ denotes the total number of the one-third octave bands.

To obtain the initial denoised speech signal, the IRM is first found. Then, the denoised speech obtained via the denoising module is fed into the modified STOI function. The following loss function is used to update the network parameters

$$\mathcal{L}(l) = \left(1 - f(S_l^{24}, \hat{S}_l^{24})\right)^2 + \lambda \left\|S_l^{24} - \hat{S}_l^{24}\right\|_F / 24 \quad (30)$$

As discussed earlier, MSE as a loss function leads to over-smoothing speech trajectories and thus generating muffled sound. Also, MSE treats each element with equal importance which is not the case in realistic audio environments. In [65], a supervised speech denoising was proposed to improve address the MSE shortcoming. The loss function used is based on the two widely used speech quality and speech intelligibility metrics, namely STOI and Perceptual Evaluation of Speech Quality (PESQ) [101].

The critical aspect regarding loss functions is their differentiability. Incorporating STOI and PESQ into the loss function is a challenging task since it cannot be differentiated using the standard algorithms such as gradient descent. A gradient approximation algorithm was introduced in the above approach whose steps are outlined below.

---

***Algorithm 2 Gradient Approximation***

**Initialization:** Noise variance $\sigma^2$, learning rate $\alpha$, initial parameters $I_0$
**for** $l = 1,2,3,\ldots,L$ **do**
    Sample $\epsilon_1, \ldots, \epsilon_n \sim N(0, I)$
    Calculate $H_i = h(w_l + \sigma\epsilon_i) - h(w_l)$ for $i = 1, \ldots, N$
    Do $w_{l+1} \leftarrow w_l - \alpha \frac{1}{N\sigma}\sum_{i=1}^N H_i / \epsilon_i$
**end for**

---

A deep neural network with perceptual connection weights was considered for monaural speech denoising in [66]. In this method, the auditory perception was incorporated into the reconstruction error to fine-tune the weights of the network. The frequency spectrum of the error was shaped in order to obtain good quality auditory masking performance. To do so, less emphasis was placed near the formant peaks, and more emphasis was placed on the spectral valleys. The following filter was considered in this method

$$P(z) = \frac{A\left(\frac{z}{\gamma_1}\right)}{A\left(\frac{z}{\gamma_2}\right)} = \frac{1 - \sum_{k=1}^q a_k \gamma_1^k z^{-k}}{1 - \sum_{k=1}^q a_k \gamma_2^k z^{-k}} \quad (31)$$

where $A(z)$ is the LPC polynomial, $a_k$s are the short-term linear prediction coefficients, $\gamma_1$ and $\gamma_2$ are the parameters controlling the energy error in the formant regions, and $q$ is the prediction error.

Then, the magnitude of the clean speech $S$ as well as the frequency response of the perceptual filter $P$ were used as a joint objective function. Mathematically, the following equation was used

$$J_{MSE} = \left\|W_f(\hat{S} - S)\right\|^2 + \left\|\hat{P} - P\right\|^2 \quad (32)$$

where the second term indicates the error between the output $\hat{P}$ and the reference $P$. This term makes use of the clean speech for fine-tuning $W_f$ as the perceptual weighting matrix which is defined as follows

$$W_f = \begin{bmatrix} P(0) & \cdots & 0 \\ \vdots & \ddots & \vdots \\ 0 & \cdots & P((N-1)\omega_0) \end{bmatrix} \quad (33)$$

with $\omega_0 = \frac{2\pi}{T}$ and $T$ being the FFT length.

To make use of the input information for a single channel speech denoising, a complex ratio masking for joint optimization of magnitude and phase was presented in [68]. The aim of this method was to derive a complex ratio mask that produces the STFT of the clean speech when applied to the STFT of the noisy speech. In other words, the following complex spectrum of the clean speech $S_{k,l}$ was considered

$$\hat{S}_{k,l}^c = \hat{M}_{k,l}^c * X_{k,l}^c \quad (34)$$

where "*" indicates complex multiplication, $\hat{M}_{k,l}^c$ is the complex IRM (cIRM), and $X_{k,l}^c$, $\hat{M}_{k,l}^c$ and $\hat{S}_{k,l}^c$ are complex numbers expressed as

$$X^c = X_r + iX_i \quad (35)$$
$$\hat{M}^c = \hat{M}_r + i\hat{M}_i \quad (36)$$
$$\hat{S}^c = \hat{S}_r + i\hat{S}_i \quad (37)$$

with the indices $r$ and $i$ denoting the real and imaginary components, respectively. Based on these definitions, Eq. (37) can be written as

$$\hat{S}_r + i\hat{S}_i = (\hat{M}_r + i\hat{M}_i) * (X_r + iX_i) \quad (38)$$
$$= (\hat{M}_r Y_r - \hat{M}_i X_i) + i(\hat{M}_r X_i + \hat{M}_i X_r)$$

The real and imaginary parts of the clean speech are then given by

$$\hat{S}_r = \hat{M}_r X_r - \hat{M}_i X_i \quad (39)$$
$$\hat{S}_i = \hat{M}_r X_i + \hat{M}_i X_r \quad (40)$$

After solving for $\widehat{M}_r$ and $\widehat{M}_i$ by using Eq. (39) and Eq. (40), the complex ideal ratio mask $\widehat{M}^c$ is computed as follows

$$\widehat{M}^c = \frac{X_r \hat{S}_r + X_i \hat{S}_i}{X_r^2 + X_i^2} + i \frac{X_r \hat{S}_i - X_i \hat{S}_r}{X_r^2 + X_i^2} \tag{41}$$

The cIRM acts similar to the Wiener filter.

In [69], a Shifted Real Spectrum (SRS) mask was proposed for a single channel speech denoising. Given a time-domain signal $x$, one can decompose it into the time-frequency domain via Discrete Time Fourier Transform (DTFT). The result is a complex spectrum $X^c$ with the real and imaginary parts of the signal denoted by $X_r$ and $X_i$, respectively.

It is well known that a time-domain signal $x$ can be represented in terms of its even and odd parts

$$x = x_{even} + x_{odd} \tag{42}$$

where $x_{even} = IDTFT(X_r)$ and $x_{odd} = IDTFT(jX_i)$. In SRS, the signal $x$ is padded with zeros to make $x(l) = 0$ when $l \leq 0$ with $l$ being the time index. The decomposition using $x_{even}$ and $x_{odd}$ parts can be stated as

$$x_{even}(l) = x_{odd}(l) = \frac{1}{2}x(l) \quad \text{if } t > 0 \tag{43}$$
$$x_{even}(l) = -x_{odd}(t) \quad \text{if } t \leq 0 \tag{44}$$

After obtaining the time-frequency representation, a time-frequency mask can be built. Based on the definition of cIRM and IRM, in SRS representations, two versions of SRS-mask can be defined: cIRM-like $cIRM_{srs}$ and IRM-like $IRM_{srs}$, in which $X_{srs}$, $V_{srs}$, and $S_{srs}$ are representations of the time-domain noisy speech $x$, noise $v$, and clean speech $s$. Using the time-frequency SRS-mask $\widehat{M}_{srs}$, the denoised speech signal can be obtained using $\hat{S} = \widehat{M}_{srs} \otimes Y_{srs}$, where $\otimes$ denotes the Kronecker product of two matrices.

A multi-objective loss function based on the parameters in PESQ was proposed in [70]. It was shown that MSE could not follow the human auditory system properly. Hence, the PESQ algorithm, which is the most widely used metric for speech quality evaluation, was adopted as a loss function. In fact, the symmetric and asymmetric disturbance terms in PESQ was incorporated into the loss function for training. These terms were computed frame-by-frame based on the clean and noisy signals.

In the LPS domain, after mean and variance normalization, the commonly used MSE loss function can be expressed as follows

$$E = \frac{1}{K} \sum_k \left( \frac{\log|S_{k,l}|^2 - \mu_k}{\sigma_k} - \frac{\log|\hat{S}_{k,l}|^2 - \mu_k}{\sigma_k} \right)^2$$
$$= \frac{1}{K} \sum_k \frac{1}{\sigma^2} \left( \log \frac{|S_{k,l}|^2}{|\hat{S}_{k,l}|^2} \right)^2 \tag{45}$$

As seen in this loss function, the loudness and threshold effects are not taken into consideration when optimizing a deep neural network.

To incorporate the perceptual features mentioned above, two disturbance terms, namely $D_t^{(s)}$ and $D_t^{(a)}$, were added to the loss function. The term $D_t^{(s)}$ considers the absolute difference between the denoised and clean loudness spectra, while the term $D_t^{(a)}$ is computed from the symmetrical disturbance but weighting positive and negative loudness differences differently. Thus, the final loss function is defined as

$$J = \frac{1}{L} \sum_l \left( E + \alpha_{D_s} D_t^{(s)} + \alpha_{D_a} D_t^{(a)} \right) \tag{46}$$

where $\alpha_{D_s}$ and $\alpha_{D_a}$ are the weighting factors and $L$ is the number of frames. Eq. (46) can be viewed as a multi-objective optimization function in which the MSE and PESQ-based disturbance terms are minimized at the same time.

An SNR-aware CNN-based single channel speech denoising was presented in [71], where it was shown that the speech denoising performance is corrupted by the mismatch of training and testing for different noise types and SNRs [55]. Several attempts have been made to classify noise types before feeding the noisy speech into a DNN. However, it would be difficult to generalize this capability due to the limitations associated with this classification problem. Noise Aware Training (NAT) has been proposed to incorporate the noise information into the input features. To cope with the noise type, two SNR-aware algorithms have been proposed allowing the denoising DNN model to achieve better speech denoising performance. First, an MTL framework is applied to approximate the clean speech signal together with the SNR level for a noisy input speech. By considering MTL, the trained CNN model is made aware of the SNR level. Second, a SNR Adaptive Denoising (SNR-AD) is introduced which includes an offline and an online mode. In the offline mode, several SNR-specific denoising models are prepared in which each model is trained by the noisy/clean pairs using the noisy speech within a particular range of SNR. Estimating the SNR is a regression problem while predicting the noise type is a classification problem. In the online mode, SNR-AD first predicts the SNR level and then selects the CNN model trained in that range.

To incorporate the ability of the SNR prediction into the denoising model, CNNs can jointly predict the primary LPS features and SNR level of the noisy input. Mathematically, the loss function is modified to incorporate the SNR estimation capability as follows

$$J = \frac{1}{N} \left[ \sum_{l=1}^{N} \|S_l - \hat{S}_l\|^2 + \lambda_{SNR} \sum_{l=1}^{N} (O_l - \hat{O}_l)^2 \right] \tag{47}$$

where $O_l$ and $\hat{O}_l$ represent the actual and estimated SNR levels of the noisy input frame at index $l$, respectively, and $\lambda_{SNR}$ is a weighting factor.

In the second algorithm, the idea of using different denoising models is exploited based on the strength of noise. Before applying the proper denoising model, a decision is made by comparing the estimated SNR to some predefined thresholds. The thresholds are specified to be

$$m = \begin{cases} 1; & \hat{O}_l > \rho_1 \\ u; & \rho_{u-1} > \hat{O}_l > \rho_u, \forall U > l > 1 \\ U; & \rho_{u-1} > \hat{O}_l \end{cases} \quad (48)$$

where $\hat{O}_l$ is the estimated SNR level, $U$ indicates the total number of denoising models, $m$ corresponds to the $m$-th denoising model, and $\rho_u$ is the $l$-th threshold.

In another MTL work in [72], a CNN-based complex spectrogram enhancement was proposed. It has been stated in [39] that by increasing the window overlap and length of the Fourier transform, the importance of phase grows. Also, as described in the joint optimization-based speech denoising methods, the phase information plays a critical rule in low SNRs. The reason for this is that signals in the Fourier domain are complex, which can be described either in terms of real-imaginary parts or magnitude-phase parts. Consider the real-imaginary parts of a noisy speech as $N_r$ and $N_i$, respectively. The phase information can be expressed as

$$\arctan \frac{X_i}{X_r} = \arctan \frac{S_i + V_i}{S_r + V_r} \quad (49)$$

In case of high SNR values, i.e. $|S_i| \gg |V_i|, |S_r| \gg |V_r|$, Eq. (49) can be simplified to

$$\arctan \frac{X_i}{X_r} = \arctan \frac{S_i}{S_r} \quad (50)$$

This indicates that the importance of the phase information grows when dealing with low SNRs.

The real-imaginary parts of the spectrogram were used as an objective function as follows

$$\Theta = \sum \|\hat{S}_i - S_i\|_2^2 + \|\hat{S}_r - S_r\|_2^2 = \sum \|\hat{S}_v - S_v\|_2^2 \quad (51)$$

where $\hat{S}_v = [\hat{S}_r \ \hat{S}_i]^T$ and $S_v = [S_r \ S_i]^T$ denote the vertically cascaded vectors of the clean and enhanced real-imaginary spectrograms, respectively. Minimizing Eq. (51) is equivalent to maximizing Segmented SNR (SSNR) [102] of the denoised signal. In this approach, it was proposed to incorporate the Log Spectral Distance (LSD) [103] into the objective function to improve speech quality. Explicitly, the LSD was modeled as an LPS reconstruction term expressed as

$$LSD = \|\log(\hat{S}_i^2 + \hat{S}_r^2) - \log(S_i^2 + S_r^2)\|_2^2 \quad (52)$$

Then, Eq. (52) was combined with Eq. (51) to form the following formula

$$\Theta = \sum \alpha_{S_v} \|\hat{S}_v - S_v\|_2^2 + \alpha_{LSD} \|\log(\hat{S}_i^2 + \hat{S}_r^2) - \log(S_i^2 + S_r^2)\|_2^2 \quad (53)$$

with the weighting terms indicated by $\alpha_{S_v}$ and $\alpha_{LSD}$. Because the first term of Eq. (53) reflects the maximization of SSNR while the second term reflects the minimization of LSD metric, this speech denoising is of the MTL type.

Since the phase information has unpredictable behavior, the deep learning models cannot learn the relationship between the clean and noisy phase. Williamson et al. [77] justified the similarity between the structures in real-imaginary spectrograms and magnitude spectrogram. Hence, instead of using the phase information directly, the spectrograms were used to compute the objective function [74]. In this approach, an extension of Eq. (53) was considered. In fact, the raw waveforms were incorporated into the objective function to improve speech quality. Mathematically, Eq. (53) was modified as follows

$$\Theta = \sum \alpha_{S_v} \|\hat{S}_v - S_v\|_2^2 + \alpha_{LSD} \|\log(\hat{S}_i^2 + \hat{S}_r^2) - \log(S_i^2 + S_r^2)\|_2^2 + \alpha_{wave} \|\hat{w}_S - w_S\|_2^2 \quad (54)$$

where $\hat{w}_S$ and $w_S$ are the clean and enhanced signals, respectively, and $\alpha_{wave}$ is a weighting parameter for the last term. The last term in Eq. (54) can be expressed using the inverse discrete Fourier transform (IDFT) as a function of $y$ and $\hat{y}$ as follows

$$\alpha_{wave} \|\hat{w}_S - w_S\|_2^2 = \alpha_{wave} \|(CU_1 \hat{S}_r - DU_2 \hat{S}_i) - (CU_1 S_r - DU_2 S_i)\|_2^2 = \alpha_{wave} \|F\hat{S}_v - FS_v\|_2^2 \quad (55)$$

where $U_1, U_2$ are the matrices used for the recovery of the even symmetry of the real part, and the odd symmetry of the imaginary part, respectively, $C$ and $D$ are the cosine and sine matirces in the IDFT, and $F$ is defined as

$$F = [CU_1 - DU_2] \quad (56)$$

Hence, the objective function is rewritten as

$$\Theta = \sum \alpha_S \|\hat{S}_v - S_v\|_2^2 + \alpha_{LSD} \|\log(\hat{S}_i^2 + \hat{S}_r^2) - \log(S_i^2 + S_r^2)\|_2^2 + \alpha_{wave} \|F\hat{S}_v - FS_v\|_2^2 \quad (57)$$

It can be seen that all parts of Eq. (57) are directly related to the output vector $S_v$.

A shortcoming of the mapping-based and masking-based methods is that they do not consider the phase mismatch problem. To address this issue, a Phase Sensitive Mask

(PSM) was proposed in [84]. Here, it is worth mentioning that the PSM values are unbounded and thus the output values of the PSM estimation are normally truncated. In this method, a joint learning algorithm was proposed to model the PSM parameters explicitly. In comparison to the conventional algorithms which implicitly learn the T-F mask, this work focused on predicting the mask by an explicit framework of its variables.

Mathematically, the following approximation of PSM was first computed under the assumption that clean speech and noise are uncorrelated

$$M_{k,l}^{PSM} \approx \sqrt{\frac{|S_{k,l}|^2}{|S_{k,l}|^2 + |V_{k,l}|^2}} \cos\theta_{k,l}^{SX} \triangleq M_{k,l}^{aPSM} \quad (58)$$

where $\theta^{SX}$ indicates the difference of the clean and noisy phase. The approximated PSM (aPSM) was described as a function of three parameters: (1) Clean speech magnitude, (2) noisy speech magnitude, (3) phase difference between clean and noisy speech spectra. A network was jointly optimized by the following loss functions

$$L_{aPSM} = \sum_{k,l}[\hat{m}_{k,l}^{aPSM}|X_{k,l}| - |S_{k,l}|\cos\theta_{k,l}^{SX}]^2 \quad (59)$$

$$L_{mag} = \sum_{k,l}[|\hat{S}_{k,l}| - |S_{k,l}|]^2 + [|\hat{V}_{k,l}| - |V_{k,l}|]^2 \quad (60)$$

$$L_{pha} = \sum_{k,l}[\cos\hat{\theta}_{k,l}^{XS} - \cos\theta_{k,l}^{XS}]^2 \quad (61)$$

In addition, since the MSE does not accurately match the human auditory perception as related to speech intelligibility, a nonlinear magnitude warping technique was considered in the estimation steps. Consequently, Eq. (60) was rewritten as

$$L_{mag}^{warp} = \sum_{t,f}[|\hat{S}_{k,l}|^\alpha - |S_{k,l}|^\alpha]^2 + [|\hat{V}_{k,l}|^\alpha - |V_{k,l}|^\alpha]^2 \quad (62)$$

To further improve performance, the sum of clean and noise estimates were minimized with respect to a given mixture, that is

$$|\hat{X}_{k,l}|^2 = |\hat{S}_{k,l}|^2 + |\hat{V}_{k,l}|^2 + 2\cos\theta_{k,l}^{SV}|\hat{S}_{k,l}||\hat{V}_{k,l}| \quad (63)$$

$$L_{add}^{warp} = \sum_{t,f}[|\hat{X}_{k,l}|^\alpha - |X_{k,l}|^\alpha]^2 \quad (64)$$

Hence, the final objective function was defined to be

$$L_{Fin} = \gamma_1 L_{aPSM} + \gamma_2 L_{mag}^{warp} + \gamma_3 L_{pha} + \gamma_4 L_{add}^{warp} \quad (65)$$

where $\gamma_1, \gamma_2, \gamma_3$ and $\gamma_4$ indicate the weight values.

## IV. COMPARISON RESULTS

In this section, experiments were conducted to compare the performance of representative methods of the speech denoising methods reviewed in this paper. These representative methods were chosen because their codes are publicly available. Four extensively used metrics of Short-Time Objective Intelligibility (STOI) [100], Perceptual Evaluation of Speech Quality (PESQ) [101], Segmented Signal-to-Noise Ratio (SSNR) [102], and Log Spectral Distance (LSD) [103] were employed to compare the methods in an objective manner. The same public domain dataset of IEEE [104] was used in the experiments. The IEEE Corpus includes 3600 speech wave (.wav) files by 20 speakers (10 females and 10 males) in which each file is about 2 seconds long. The speakers are from two regions of the Pacific Northwest (PN) and the Northern Cities (NC) reading the IEEE "Harvard" sentences. Noisy dataset was made by adding two types of noises (speech type noise (Babble) and non-speech type noise (Factory) noise signals) to the clean dataset. Three levels of SNR were made as -2, 0, and 5 dB. Both noise types were used for training, evaluation, and testing parts. The LPS features were considered as the input using 256-point FFT and 50% overlap. The dataset was randomly partitioned into three parts with no overlap: 70% for training, 20% for testing, and 10% for validation.

To provide a comprehensive comparison, the conventional methods of Wiener filtering [25] and IBM [83], the single objective deep learning methods of DNN-IBM [83], DNN-IRM [87], the hybrid method of DNN-Hybrid [58], and the multi-objective method of DNN-MultiObjective (DNN-MO) [74] were examined considering the availability of their codes in the public domain.

Table I through Table III exhibit the performance of the denoised signals in terms of the above four metrics for three SNR levels. As can be seen from these tables, the DNN-Hybrid and DNN-MO methods provided superior outcomes. This was expected since they utilize more information to optimize a deep neural network. Another key observation from these tables is that in lower SNR levels, the multi-objective method performed better than the DNN-Hybrid method. Basically, as the SNR level becomes higher, the clean speech characteristics can be obtained well using the conventional methods. In other words, at higher SNRs, the difference between the DNN-Hybrid and DNN-MO method becomes small or rather negligible.

In general, since the multi-objective deep learning methods use richer information to achieve speech denoising, their performance in comparison to the other methods is better. In fact, what plays a critical role in achieving better performance is the terms selected for the optimization. Another important issue that is worth noting here is that environmental noises cannot be modeled accurately by the conventional methods. However, the deep learning methods can model the nonlinear relationship between noisy speech and clean speech signals more effectively and thus their performance in general is better even in not previously encountered noisy environments.

## V. POSSIBLE FUTURE DIRECTIONS

It is important to note that all the above deep learning methods are supervised methods, meaning that clean speech signals are assumed to be available for their training. In these methods, clean speech signals are used as the target. In practice, speech sentences spoken in the field differ from database speech sentences that are used for training. As a result, training on one set of speech signals does not guarantee the same performance when a different set of speech signals are spoken in the same noise environments. In real-world scenarios, clean speech signals are in fact not available. In [105], a speech denoising method was introduced which does not rely on the availability of clean speech signals to achieve training and it uses the mixture for both the input and the target of a deep neural network. This method eases the major assumption in the existing speech denoising methods and allows the training process to be conducted in an online or self-supervised manner without having access to clean speech signals. We envision that much progress can be made by developing different variations of the method described in [105].

Also, in the existing speech denoising methods, all noises are treated the same. We envision another future improvement can be achieved by designing deep neural networks for different types of noise and thus using a bank of deep neural networks depending on the noise characteristics encountered in the field. Since in practice or real-world scenarios, there are many different noise types, unsupervised noise classification approaches such as the one in [106] can be used to train these deep neural networks in an online manner depending on the noise type encountered in the field.

TABLE I. Average performance metrics of representative speech denoising methods in non-speech type noise (Factory – $N_1$) and speech type noise (Babble – $N_2$) at -2 dB SNR.

| Method \ Metric | STOI ($N_1$) | STOI ($N_2$) | PESQ ($N_1$) | PESQ ($N_2$) | LSD ($N_1$) | LSD ($N_2$) | SSNR ($N_1$) | SSNR ($N_2$) |
|---|---|---|---|---|---|---|---|---|
| Noisy Signal | 0.6235 | 0.6086 | 1.3057 | 1.3114 | 2.3968 | 1.9121 | -4.4675 | -3.8066 |
| Wiener | 0.5929 | 0.5450 | 1.4301 | 1.2305 | 1.5489 | 1.4658 | 0.2607 | -0.6993 |
| IBM | 0.6894 | 0.6741 | 1.3694 | 1.3354 | 1.7839 | 1.5469 | 1.0894 | 1.1152 |
| DNN – IBM | 0.7366 | 0.7119 | 1.4209 | 1.3737 | 1.3953 | 1.2860 | 0.4422 | 0.6157 |
| DNN – IRM | 0.7577 | 0.7283 | 1.6558 | 1.4961 | 1.0895 | 1.0649 | 0.5055 | 0.3867 |
| DNN – Hybrid | 0.8010 | 0.7542 | 1.6850 | 1.5813 | 0.9157 | 0.8812 | 0.6024 | 0.7312 |
| DNN – MO | 0.8159 | 0.7614 | 1.7068 | 1.6573 | 0.8556 | 0.8331 | 0.8011 | 0.7556 |

TABLE II. Average performance metrics of representative speech denoising methods in non-speech type type (Factory – N1) and speech type noise (Babble – N2) at 0 dB SNR.

| Method \ Metric | STOI ($N_1$) | STOI ($N_2$) | PESQ ($N_1$) | PESQ ($N_2$) | LSD ($N_1$) | LSD ($N_2$) | SSNR ($N_1$) | SSNR ($N_2$) |
|---|---|---|---|---|---|---|---|---|
| Noisy Signal | 0.6689 | 0.6585 | 1.3607 | 1.3689 | 2.3088 | 1.8318 | -3.6128 | -2.9568 |
| Wiener Filter | 0.6869 | 0.6040 | 1.5274 | 1.3041 | 1.4481 | 1.3931 | 0.8107 | -0.1013 |
| IBM | 0.7031 | 0.6968 | 1.4052 | 1.4012 | 1.7642 | 1.5125 | 0.5024 | 0.6587 |
| DNN – IBM | 0.7831 | 0.7567 | 1.5740 | 1.4368 | 1.3538 | 1.3125 | 2.0030 | 1.8394 |
| DNN – IRM | 0.7991 | 0.7758 | 1.8233 | 1.6141 | 0.9668 | 0.9709 | 1.5943 | 1.5159 |
| DNN – Hybrid | 0.8312 | 0.8173 | 1.9547 | 1.8333 | 0.8123 | 0.8206 | 2.1415 | 1.8012 |
| DNN – MO | 0.8450 | 0.8234 | 2.1005 | 1.9673 | 0.7315 | 0.7519 | 1.9614 | 1.9684 |

TABLE III. Average performance metrics of representative speech denoising methods in non-speech type noise (Factory – N1) and speech type noise (Babble – N2) at 5 dB SNR.

| Method \ Metric | STOI ($N_1$) | STOI ($N_2$) | PESQ ($N_1$) | PESQ ($N_2$) | LSD ($N_1$) | LSD ($N_2$) | SSNR ($N_1$) | SSNR ($N_2$) |
|---|---|---|---|---|---|---|---|---|
| Noisy Signal | 0.7837 | 0.7788 | 1.5424 | 1.5504 | 2.0065 | 1.5628 | -0.6908 | -0.0727 |
| Wiener Filter | 0.7878 | 0.7849 | 1.8624 | 1.5558 | 1.3534 | 1.2688 | 2.4677 | 1.8805 |
| IBM | 0.8046 | 0.7975 | 1.7056 | 1.6652 | 1.5193 | 1.4658 | 2.1259 | 2.0126 |
| DNN – IBM | 0.8657 | 0.8457 | 1.9607 | 1.7420 | 1.4568 | 1.3780 | 5.2782 | 5.0936 |
| DNN – IRM | 0.8841 | 0.8627 | 2.2386 | 2.0020 | 0.8084 | 0.7725 | 4.4368 | 4.2259 |
| DNN – Hybrid | 0.9111 | 0.8827 | 2.3071 | 2.1101 | 0.7130 | 0.6852 | 5.5115 | 5.3230 |
| DNN – MO | 0.9362 | 0.9135 | 2.5403 | 2.3114 | 0.6823 | 0.6533 | 6.4371 | 6.1123 |

## VI. CONCLUSION

A review of multi-objective deep learning speech denoising methods has been covered in this paper. These recent methods denote the current state-of-the-art in speech denoising. To set the stage for this review, an overview of conventional, single objective deep learning, and hybrid methods was first presented. This overview was followed by a review of the mathematical framework of the existing multi-objective deep learning methods. Representative methods in the speech denoising categories whose codes are publicly available were then compared by considering the same public domain dataset and four widely used objective metrics. The comparison conducted has shown the effectiveness of the multi-objective deep learning methods, in particular in low SNRs as compared with the other existing methods.